# A general illumination method to predict bifacial photovoltaic system performance

Erin M. Tonita,[1,*,†] Christopher E. Valdivia,[1] Annie C. J. Russell,[1] Michael Martinez-Szewczyk,[2] Mariana I. Bertoni,[2] and Karin Hinzer[1]

[1]SUNLAB, Centre for Research in Photonics, University of Ottawa, Ottawa, ON, K1N6N5, Canada

[2]DEfECT Lab, Arizona State University, Tempe, AZ, 85287-9709, USA

*Correspondence: etoni044@uottawa.ca

†Lead contact

Erin Tonita is currently a PhD candidate in Physics at the University of Ottawa under the supervision of Dr. Karin Hinzer of the SUNLAB. Her research activities focus on the modelling and characterization of bifacial photovoltaic technologies for diesel-displacement in the Arctic. Her hobbies include biking, interpretive dance, and astrophotonics.

Dr. Christopher Valdivia is a Senior Research Scientist in the University of Ottawa's SUNLAB, with >20 years of research in photonics. He holds a PhD from the Optoelectronics Research Centre, University of Southampton, UK (2007). His research focus is renewable solar energy generation and photonic power, with activities spanning novel photonic device design to systems energy yield modeling.

Annie Russell is a PhD candidate at the University of Ottawa SUNLAB where she develops the bifacial PV performance software, DUET. She works from unceded territory of the Musqueam, Squamish, and Tsleil-Waututh Nations.

Michael Martinez-Szewczyk is currently a PhD student at Arizona State University under the supervision of Dr. Mariana Bertoni. His research activities focus on novel low-resistance silver metallization for low-temperature applications and their use in next generation photovoltaic devices by way of dispense printing.

Mariana I. Bertoni is a Professor at Arizona State University. She received her PhD in 2007 from Northwestern University, followed by a postdoctoral fellowship at the Massachusetts Institute of Technology. Her group focuses on defect engineering and advanced characterization of solar cells and modules.

Karin Hinzer is Vice-Dean, Research of the Faculty of Engineering and a Professor at the School of Electrical Engineering and Computer Science with a cross-appointment in the department of Physics at the University of Ottawa, and the University Research Chair in Photonic Devices for Energy. She has published over 190 refereed papers, trained over 170 highly-qualified personnel and her laboratory has spun-off three Canadian companies in the energy sector. Her research interests include new materials, high efficiency light sources and light detectors, photovoltaics, solar modules, new electrical grid architectures and power converters.

## SUMMARY

Bifacial photovoltaic technologies are estimated to supply >16% of global energy demand by 2050 to achieve net-zero greenhouse gas emissions. However, the current IEC bifacial measurement standard (IEC 60904-1-2) does not provide a pathway to account for the critical effects of spectral or broadband albedo on the rear-side irradiance, with in-lab characterization of bifacial devices limited by overestimation of rear incident irradiance, neglecting spectral albedo effects on the rear, or both. As a result, prior reports have limited applicability to the diverse landscapes of bifacial photovoltaic deployments. In this paper, we identify a general bifacial illumination method which accounts for spectral albedo while representing realistic system operating conditions, referred to as the *scaled rear irradiance (SRI) method*. We describe how the SRI method extends the IEC standard, facilitating indoor testing of cell or module performance under varied albedo with standard solar simulator set-ups. This enables improved comparisons of bifacial technologies, application-specific optimization, and the standardization of bifacial module power ratings.

## INTRODUCTION

The International Energy Agency has outlined a roadmap to achieving net-zero greenhouse gas emissions by 2050, requiring a rapid transformation of the global energy sector towards renewable technologies.[1] While the cumulative total capacity of deployed photovoltaics (PV) exceeded 1 TW in 2022, annual PV deployments on the scale of several terawatts by the mid-2030s are anticipated.[2] By 2050, PV technologies are projected to supply 20% of global energy and >70% of global electricity.[1] Of this PV, bifacial technologies are expected to dominate the market share by >80%[3] due to the bifacial benefits of increased power production per area, increased module lifetime, and cost-competitive manufacturing.[4] Planning of system deployments to meet global energy demand requires accurate and standardized measurements of bifacial PV. Standard measurements are quoted on commercial module datasheets and used to plan system performance needs, which impacts system cost, technology choice, material consumption, land use, and grid stability.

In recent years, progress has been made on standardizing bifacial measurement procedures.[5-8] In 2019, the International Electrotechnical Commission (IEC) published a technical specification, IEC 60904-1-2,[5] which defines two standard methods of characterizing bifacial PV devices where the front-face is illuminated with 1000 W/m$^2$ while the rear-face is illuminated in the range of 100-250 W/m$^2$. Work is underway to evaluate the merits and drawbacks of these two measurement methods.[6,8]

Rear-side incident irradiance contributions can vary widely from 0 W/m$^2$ to 700 W/m$^2$ depending on the albedo of the surrounding environment, orientation of the module, shading, angle of the sun, atmospheric conditions, and temporal fluctuations.[7,9] In a recent systems-level study, Onno *et al*.[9] evaluated incident photon-flux on bifacial modules, finding rear irradiances from 0-400 W/m$^2$ under typical operating conditions, with the two most important factors governing irradiance variation being tracking type and the ground albedo. Ground albedo changes naturally across diverse global environments and can be engineered to increase energy yield. An irradiance of 200 W/m$^2$ is the value selected by IEC to represent a typical rear side contribution for bifacial testing methods in IEC 60904-1-2. However, this rear irradiance is not attributed to any particular ground condition nor does it account for spectral variation on the rear face, as it applies the standard air mass (AM) spectrum of AM1.5G to both faces. In the field, rear incident light has a unique spectral shape determined by the physical properties of the ground cover, which previous research demonstrates in some cases to yield device power variation of >5%.[10,11]

In this paper, we propose a method to complement IEC measurement procedures which accommodates the effects of spectral albedo and realistic rear-front irradiance ratios. We model the performance of a typical bifacial silicon heterojunction (SHJ) solar cell under three bifacial illumination techniques: the two standard methods outlined by IEC, and our new spectral albedo approach, the *scaled rear irradiance* method. Finally, we demonstrate at the cell/module level the ability of our proposed method to represent and predict energy yield by comparison to a detailed system-level optical and electrical performance model.

## BIFACIAL ILLUMINATION METHODS

### IEC 60904-1-2

In the IEC *bifacial* method, the 1-sun (1000 W/m$^2$) AM1.5G spectrum illuminates the front side, while the same spectrum is scaled to a value between 0.1-0.25 suns (100-250 W/m$^2$) on the rear side. In the IEC *equivalent irradiance (EQI)* method, 1-sun front side irradiance, $G_{f,\text{EQI}}(\lambda)$, is increased according to the device's bifaciality, $\varphi$. For example, for a rear irradiance of 0.2 suns, the applied spectrum is:

$$G_{f,\text{EQI}}(\lambda) = AM1.5G(\lambda) + [0.2\ \varphi\ AM1.5G(\lambda)] \qquad \text{(Eq. 1)}$$

with wavelength given by $\lambda$. For the SHJ device examined in this paper, the bifaciality factor is 0.96, but can range from 0.6 to 1 depending on the cell and module technology. Technologies with lower bifaciality will have lower bifacial gain than what is presented in this paper, but will exhibit similar trends. For example, in our device, to mimic bifacial illumination with a rear incident contribution of 200 W/m$^2$, the equivalent front side illumination is 1192 W/m$^2$. The EQI method is intended to provide a simpler measurement procedure, eliminating the need for customized device mounting, or additional lamps and optics.

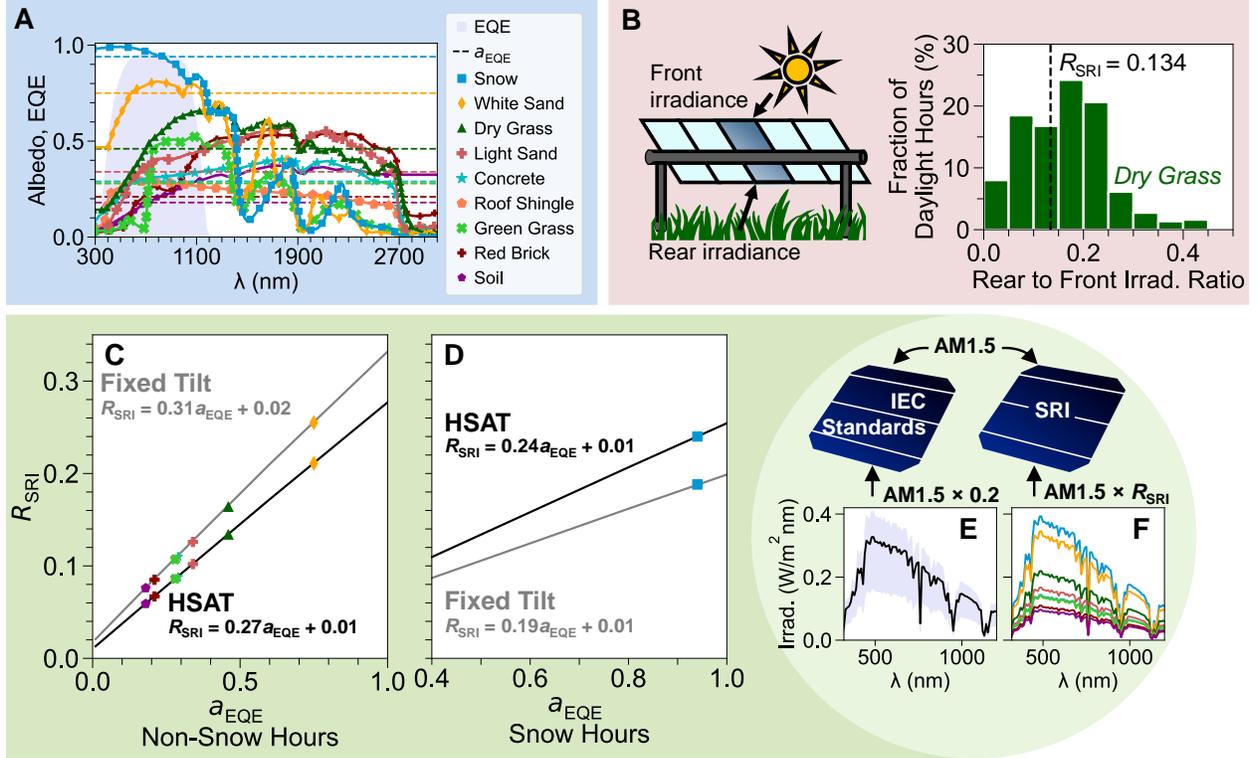

**Figure 1. The Scaled Rear Irradiance Method**

(A) Spectral albedos considered in this work, digitized from Russell et al.[11] and Gueymard et al.[12] Rear-side EQE, as measured by the authors for our SHJ device, is provided in the background. Dashed lines indicate the EQE-weighted albedo, as defined by Equation 5.

(B) Depiction of the SRI method calibration for dry grass. The histogram displays the fraction of daylight hours which occur at a particular ratio of rear to front plane-of-array irradiance on a module in a HSAT array.

(C and D) The relationship between systems-input spectrally weighted albedo and $R_{SRI}$ for fixed-tilt and HSAT tracking types during (C) non-snow hours and (D) snow hours.

(E and F) Illustration of the irradiance applied under (E) the IEC bifacial method compared to (F) the spectral albedo-determined irradiance of the SRI method.

## The SRI Method

In our proposed *scaled rear irradiance (SRI)* method, the rear-side irradiance is determined by the spectral albedo of the ground. The rear incident spectrum, $G_{r,SRI}(\lambda)$, is:

$$G_{r,SRI}(\lambda) = AM1.5G(\lambda)\,R_{SRI}(x, \tau) \qquad \text{(Eq. 2)}$$

where $R_{SRI}$ represents a system-level ratio of rear-to-front module plane of array irradiance (PoA) for spectral albedo $x$ and system tracking type $\tau$. In this paper, $x$ represents the ground types studied: snow, white sand, dry grass, light sand, concrete, roof shingle, green grass, red brick, and soil. Spectral albedo data is retrieved from *Russell et al.*,[11] with light sand and soil provided by the SMARTS database.[12] Data spanning the wavelength range of 280-3000 nm is digitized and reproduced in Figure 1A. We consider tracking types of horizontal single-axis tracked (HSAT) and latitude fixed-tilt. Values of $R_{SRI}$ are given in Table 1, while their calibration is discussed in the following sections.

## The SERI Method

Analogous to the IEC equivalent irradiance method, in the *scaled equivalent rear irradiance (SERI)* method the rear-side irradiance is added to the front:

$$G_{f,SERI}(\lambda) = AM1.5G(\lambda) + [R_{SRI}\,\varphi\,AM1.5G(\lambda)] \qquad \text{(Eq. 3)}$$

Due to the equivalency of this method and the SRI method for optically linear device technologies, as is the case for most PV technologies, all results presented in this work are under illumination on both front and rear faces.

**Table 1. Summary of Albedo and SRI Method Parameters**

| Ground cover, $x$ | Soil | Red brick | Green grass | Roof shingle | Concrete | Light sand | Dry grass | White sand | Snow |
|---|---|---|---|---|---|---|---|---|---|
| |  |  |  |  |  |  |  |  |  |
| Broadband albedo, $a_{bb}$ | 0.19 | 0.23 | 0.24 | 0.26 | 0.29 | 0.33 | 0.44 | 0.67 | 0.85 |
| EQE-weighted albedo, $a_{EQE}$ | 0.18 | 0.21 | 0.28 | 0.28 | 0.29 | 0.34 | 0.46 | 0.75 | 0.94 |
| Horizontal single-axis tracked irradiance ratio, $R_{SRI}$ | 0.059 | 0.067 | 0.086 | 0.086 | 0.088 | 0.102 | 0.134 | 0.211 | 0.240 |
| Fixed-tilt irradiance ratio, $R_{SRI}$ | 0.076 | 0.085 | 0.107 | 0.107 | 0.110 | 0.126 | 0.164 | 0.255 | 0.188 |

## Calibration of the SRI Method

As the two most important factors governing rear-side irradiance are albedo and tracking-type,[9] $R_{SRI}$ calibration must occur for each system configuration of interest. We exemplify our approach using a bifacial horizontal single-axis tracked array located in Boulder, Colorado. Results for a latitude fixed-tilt system are additionally summarized in Table 1 and Figure 1.

Calibration of the SRI method requires a bifacial PV performance model that inputs broadband albedo and outputs front and rear insolation. *Broadband albedo*, $a_{bb}$, is calculated by weighting the spectral albedo, $A_x(\lambda)$, by the AM1.5G spectrum:[13]

$$a_{bb} = \frac{\int A_x(\lambda)\, AM1.5G(\lambda)\, d\lambda}{\int AM1.5G(\lambda)\, d\lambda} \qquad \text{(Eq. 4)}$$

Integration bounds are determined by the pyranometer sensitivity, in our case 280-3000 nm. We performed this calibration using the detailed system-level 3D view factor model, DUET.[14] DUET applies system array configurations and environmental conditions to calculate hourly front and rear irradiance profiles and *I-V* curves, among other outputs. This model has been validated against field data and other software.[14]

Typical meteorological year (TMY) input data was obtained from NREL's National Solar Radiation Database (NSRDB).[15] The chosen representative bifacial PV array consists of 72-cell modules with a 1-in-portrait configuration and a 1.22 m ground clearance. Input single-diode model parameters are taken from our optoelectronic cell model for the encapsulated bifacial SHJ structure detailed in Tonita et al.[16] For a summary of all system inputs, see the Supplemental Information Table S2.

System performance models typically apply broadband albedo value(s) to reduce computational cost, so the effects of spectrally-resolved albedo are introduced by the application of an external quantum efficiency (EQE)-weighted albedo, $a_{EQE}$:

$$a_{EQE} = \frac{\int A_x(\lambda)\, EQE(\lambda)\, AM1.5G(\lambda)\, \lambda\, d\lambda}{\int EQE(\lambda)\, AM1.5G(\lambda)\, \lambda\, d\lambda} \qquad \text{(Eq. 5)}$$

This approach reduces output electrical power discrepancy caused by spectral versus broadband albedo use, as it weights the spectral albedo according to the responsivity of the specific technology.[13] These EQE-weighted albedos are provided in Table 1 and visualized in Figure 1A. Broadband albedo values are provided in Table 1 for comparison.

Finally, to determine the representative rear-to-front incident irradiance ratio for each spectral albedo ($R_{SRI}$ in Equation 2 and 3), we calculate hourly 2D front and rear irradiance profiles on a representative module in the bi-HSAT array over the course of a year. Hourly albedo data is categorized into two datasets: hours with snow and hours without snow. To identify the representative ratio for snowy albedo, the albedo of snow is applied to only snowy hours (83 days). For all other albedos, they are applied to non-snowy hours (282 days).

The ratio of rear (r) to front (f) module plane of array (PoA) insolation across all timesteps $t$ with albedo $x$ determines $R_{SRI}$:

$$R_{SRI} = \frac{\sum_t PoA_r(t)}{\sum_t PoA_f(t)} \qquad \text{(Eq. 6)}$$

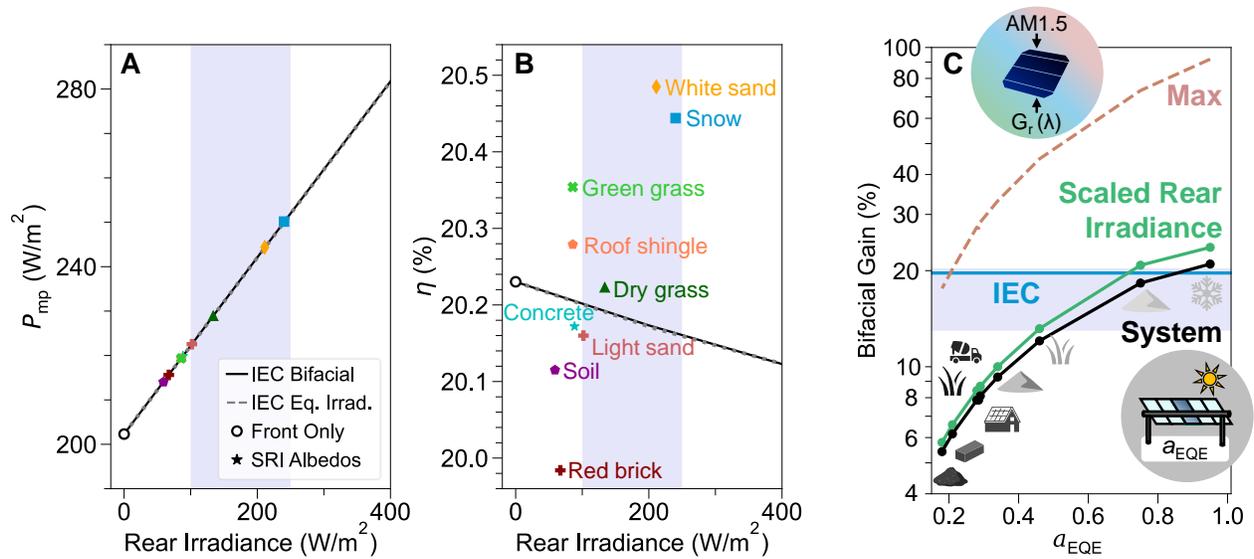

**Figure 2. Performance of Bifacial Illumination Methods on a Device and System Level**

(A) The maximum power produced by a SHJ cell using the IEC and SRI methods. Front-only results are given by the white circles, while the shaded background corresponds to the recommended IEC testing range.

(B) The device bifacial efficiency for these methods. SRI efficiency is calculated using Equations 7 and 8, resulting in the deviation from the AM1.5G IEC lines.

(C) Bifacial gain calculated using two methods. In black, the bifacial energy yield gain of a bi-HSAT module using a validated system-level 3D view factor model, DUET. The remaining lines are calculated as the gain in device power under bifacial illumination conditions compared to front-only illumination for device measurement methods. The shaded background around the IEC 60904-1-2 line corresponds to the rear irradiance range of 100-250 W/m². The "Max" line represents the maximum theoretical gain due to ground-reflected light in the ideal case devoid of any ground shading. Data is spaced according to the EQE-weighted broadband albedo ($a_{EQE}$) input into DUET, with icons depicting the ground condition associated with each data point.

The distribution of $R_{SRI}$ across timesteps for dry grass is provided as an example in Figure 1B, while all identified irradiance ratios are provided in Figure 1C, 1D, and summarized in Table 1. Figure 1C depicts the relationship between system-input albedo, $a_{EQE}$, and $R_{SRI}$ for fixed tilt and HSAT array configurations during non-snowy hours, while the relationship for snowy hours is given in 1D. It is possible to extract from these subplots the appropriate $R_{SRI}$ value for any albedo of interest.

Figure 1E depicts the rear-side irradiance applied during the IEC bifacial method, while 1F shows the irradiance used in the SRI method using calibrated $R_{SRI}$.

## RESULTS

### Device-Level Performance

We simulated simultaneous front and rear illumination in Synopsys TCAD Sentaurus using an optoelectronic model for textured crystalline silicon (c-Si) based encapsulated bifacial SHJ solar cells.[16] A schematic of our cell structure under bifacial illumination scenarios is shown in Figure S1.

To align with standard test conditions, cells were maintained at 25°C and incoming light was normally-incident for all calculations. For a detailed description of our model and model validation see the Supplemental Information and Tonita et al.[16]

The maximum power and efficiency output under the SRI method are provided in Figure 2A and 2B, with IEC standards given by dotted and solid lines for comparison. Front-only illumination is represented by an open circle at 0 W/m² rear irradiance. For the IEC methods, we extend the recommended testing range of rear-side irradiances of 100-250 W/m² to 0-400 W/m² and indicate the recommended testing regime by the shaded backgrounds throughout Figure 2.

The conventional monofacial definition of efficiency ($\eta$) is a ratio between total electrical output power, $P_{mp}$, and total incident power, $P_{tot}$. As the definition of efficiency under bifacial illumination is not yet clearly defined,[7] we consider the total incident power to be the sum of front, $P_{inc,f}$, and rear, $P_{inc,r}$, incident power contributions. Thus bifacial efficiency becomes:

$$\eta = \frac{P_{mp}}{P_{tot}} = \frac{P_{mp}}{P_{inc,f} + P_{inc,r}} \qquad \text{(Eq. 7)}$$

To include the effects of spectral shape on efficiency in the SRI method, $P_{inc,r}$ must be set to:

$$P_{inc,r} = 1000 \frac{W}{m^2} R_{SRI} \frac{a_{bb}}{a_{EQE}} \qquad \text{(Eq. 8)}$$

This accounts for the scaling of the AM1.5G rear irradiance according to the spectral shape of the ground albedo. For example, relatively more photons are present in the c-Si absorption range when the spectral albedo is snow compared to AM1.5G, resulting in $a_{EQE} > a_{bb}$. To achieve the same $P_{mp}$ using a spectrum modified by the ground spectral albedo, less total irradiance must be applied to the rear, given by Equation 8 above. This definition is equivalent to integrating the rear incident spectrum reflected off the ground, which emphasizes the power conversion efficiency of all light incident on the device and is necessary for calculations of thermalization and heating.

Figures 2A and 2B show significant variation in $P_{mp}$ and efficiency depending on the applied rear irradiance. The two IEC methods have near-overlapping lines in all subplots, demonstrating a negligible difference between the equivalent irradiance method and the bifacial method, as designed. In Figure 2B, their efficiencies decrease with increased rear contribution as rear-injected carriers require on-average longer diffusion lengths to reach the front p-n junction, resulting in higher recombination loss. The efficiency becomes a weighted average of the front and rear efficiency, with the weight determined by their relative irradiance contributions.

The SRI method is intended to provide a method to test the effects of spectral albedo on device performance. The rear irradiances resulting from $R_{SRI}$ for each ground cover fall within the typical rear-irradiance operating range reported at the system-level,[8] with input incident AM1.5G rear irradiance between 59-240 W/m$^2$ on the x-axes of Figures 2A and 2B. The maximum and minimum output power using the SRI method for the considered albedo dataset covers a range of values spanning 36 W/m$^2$.

As efficiency includes the effects of spectral shape through Equation 8, we calculate efficiency differences between +0.33% abs. to -0.20% abs. compared to IEC methods that assume AM1.5G on both faces. For example, snow reflects >90% of light in the c-Si absorption range, leading to an efficiency 0.30% abs. higher than what is given by the IEC methods. Thus, the efficiency of a device at absorbing incident radiation under varying ground conditions cannot be accurately characterized without considering changes in spectral shape.

## Energy Yield Prediction

The SRI method provides illumination conditions that match operating conditions for a given combination of ground cover and system design, more closely representing bifacial system-level performance than 200 W/m$^2$ IEC rear irradiance. To illustrate, we calculate the energy yield for a module in the bi-HSAT array located in Boulder, Colorado using DUET. The black line of Figure 2C displays its bifacial gain, defined as the percentage increase in energy yield compared to an equivalent monofacial module. Different ground conditions are plotted on the x-axis, with spacing determined by the EQE-weighted albedo applied in DUET.

For comparison, device-level characterization methods are plotted using bifacial $P_{mp}$ gain, defined as the gain in power under bifacial illumination conditions compared to front-only AM1.5G illumination. This is calculated for the IEC bifacial and SRI methods. As an additional comparison point, the maximum bifacial $P_{mp}$ gain due to ground-reflected light is plotted in red, for the case where $R_{SRI} = a_{EQE}$ in Equation 2.

Since the system-level ratio of rear-to-front total annual irradiance dictates the rear irradiance applied in the SRI method, the resulting bifacial gain tracks the full system-level analysis well, with only a 1.7% abs. discrepancy with DUET on average across varied ground conditions. The agreement between bifacial $P_{mp}$ gain and DUET energy yield gain confirm that in-lab device characterization using the SRI method is sufficient to capture the main sources of energy yield variation – namely albedo and tracking type. While $R_{SRI}$ was calibrated for Boulder, Colorado (40°N, diffuse fraction 0.36), energy yield gains predicted using $R_{SRI}$ are within 2.0% and 0.7% abs. on average in Phoenix, Arizona (33°N, diffuse fraction 0.27) and Ottawa, Canada (45°N, diffuse fraction 0.43), respectively. Thus, it is possible to use the values presented in this paper for other locations, with an $R_{\mathrm{SRI}}$ error of < 7%. Further details on the impact of geography are provided in the Supplemental Information and Table S1.

As a contrast, IEC standards do not provide a pathway for adjusting rear irradiance given system configuration and ground cover, and thus produce a flat bifacial gain in Figure 2C. For a rear irradiance of 200 W/m$^2$, the IEC method over-estimates bifacial gain by between +14% abs. to +1.4% abs for all ground covers except snow. This discrepancy reinforces the need for the SRI method to realistically represent a range of operation conditions.

## DISCUSSION

### The SRI Method as a Full Spectrum Calculation

With the development of LED solar simulator technologies and the possibility for use of custom spectral filters, bifacial measurements with unique front and rear spectra could be implemented in the future. Although the SRI method utilizes AM1.5G on the front and the rear of cells or modules, it is possible to implement this method with rear-side spectral shape given by the spectral albedo, $A_x(\lambda)$.

Rather than a scaling of AM1.5G, the rear spectrum, $G_{r,\mathrm{SRI}}(\lambda)$, is applied as given:

$$G_{r,\mathrm{SRI}}(\lambda) = AM1.5G(\lambda) \; \frac{A_x(\lambda)}{a_{\mathrm{bb}}} \; R_{\mathrm{SRI}}(x, \tau) \qquad \text{(Eq. 9)}$$

The device power output by Equation 9 is closely matched by the SRI method depicted in Figure 2A, as the effect of spectral shape is included in the calibration of the SRI method through $a_{\mathrm{EQE}}$. This method can additionally be applied to any solar spectrum of interest, not just AM1.5G, such as spectra with varying air mass, precipitable water, or aerosol optical depth. Further details are provided in Supplemental Information.

### Impact and Significance

In this work, *the SRI method* was described and shown to be an improved method for measuring and modelling bifacial devices that is representative of annual performance, particularly due to its ability to represent different surface albedos. Incorporating this new method into future bifacial standards would provide a consistent methodology for testing bifacial devices with spectral or broadband albedo, corresponding to globally-varying illumination conditions. This is of particular importance as renewable energy penetration increases towards a net-zero world, with bifacial PV projected to contribute ~16% of the global energy supply by 2050, around 30,000 TWh annually.[1,3]

Early implementation of this method into IEC standards can:

1. Enable comparisons between existing and emerging bifacial technologies;

For example, the SRI method can provide a consistent approach for future experimental studies of emerging tandem solar devices, which have a strong spectral dependence due to their design of segmenting absorption into series-connected top and bottom cells. Tandem devices are anticipated to be among the next-generation of silicon solar cell technologies due to their ability to exceed the single-junction efficiency limit.[9]

2. Enable application-specific optimization;

With the SRI method, it is possible to optimize rear-side passivation and anti-reflection coatings for specific illumination conditions. For example, in the case of silicon heterojunction cells with UV-absorptive ground conditions such as green grass, tolerances on rear-side ITO properties are broadened and thicker layers can be fabricated.

3. Increase PV deployments in non-traditional markets;

This method also highlights the favorable conditions of snow accumulation present in high-latitude locations by enabling in-lab device characterization under snowy ground conditions.

4. Improve system deployment sizing;

Adopting this method can also reduce investment risk in PV deployments, and impact system planning. For example, a 1% rel. increase in efficiency of the bi-HSAT system located in Boulder, Colorado results in a 1.1% increase in annual energy yield. More-accurate measurement capabilities can affect future terawatt deployments on the several gigawatt scale, therefore changing system cost, material consumption, and land use. Accurate system predictions and planning can also potentially reduce system inefficiencies caused by mismatch in PV generation and grid load, or impact the sizing of storage technologies.

5. Support bifacial power ratings.

Furthermore, as instantaneous front-only AM1.5G illumination power ratings are currently used for monofacial and bifacial modules alike, this method can inform future bifacial power ratings. We suggest manufacturers provide bifacial power ratings for different common ground conditions and tracking types, such as snow, sand, grass, and soil for single-axis tracked systems and fixed-tilt.

Estimates of the bifacial power produced under the SRI method from existing 1-sun front-only module power ratings can be done given the linear short-circuit current ($J_{sc}$) and logarithmic open-circuit voltage ($V_{oc}$) relationship with irradiance. For instance, assuming a negligible change in fill factor, the maximum power produced under bifacial illumination with the SRI method, $P_{mp,SRI}$, can be calculated as follows:

$$P_{mp,SRI} = P_{mp,1-sun} X \left(1 + \frac{0.2569\,V}{V_{oc}} \ln(X)\right) \qquad \text{(Eq. 10)}$$

where $X$ is the effective increase in incident irradiance for a particular albedo:

$$X = (1 + R_{SRI})\,\varphi \qquad \text{(Eq. 11)}$$

The use of bifacial power ratings defined with the SRI method will highlight the benefit of bifacial technologies, while facilitating PV technology choice that is best suited for the needs of individual projects.

## Conclusion

In this study, we described a general bifacial illumination method, *the scaled rear irradiance method,* which can predict outdoor system performance under varying ground conditions by appropriate scaling of the standard AM1.5G spectrum. We outline how the SRI method can be used as an extension to IEC 60904-1-2 bifacial measurement standards to (1) capture efficiency differences under varied spectral albedo and (2) represent system-specific illumination levels through in-lab bifacial device measurements. Implementation of this approach in IEC standards would provide commercial bifacial module manufacturers with a power ratings methodology for common ground covers, such as snow, sand, grass, and soil. This additional rating would allow for technology comparisons and design optimizations specific to the planned PV deployment, while potentially hastening bifacial PV adoption. As bifacial PV deployments are already exponentially rising each year[2] and module lifetimes exceed 25 years, improvements to bifacial measurement standards must happen rapidly to keep pace with increasingly varied installation conditions, reduce financial risk, maximize system energy yields, and limit greenhouse gas emissions.

## DATA AND CODE AVAILABILITY

The data supporting the findings of this study are available in the main text and Supplemental Information and are additionally available upon request from the corresponding author. Details regarding the code used in this paper are published in Russell et al.[14] and Tonita et al.[16]

## SUPPLEMENTAL INFORMATION

Supplemental Information is attached to this document.

## ACKNOWLEDGEMENTS

The authors thank CMC Microsystems and the Natural Sciences and Engineering Research Council of Canada [NSERC CREATE 497981, NSERC STPGP 521894, NSERC CGS-D, NSERC RGPIN-2015-04782] for their funding support. MB and MMS acknowledge funding from the Department of Energy (DOE) under contract No. DE-EE0008166. Any opinions, findings, and conclusions or recommendations expressed in this material are those of the author(s) and do not necessarily reflect the views of the Department of Energy.

The University of Ottawa is on the unceded territory of the Anishinaabe Algonquin Nation.

## AUTHOR CONTRIBUTIONS

Conceptualization, E.M.T., C.E.V., M.I.B, and K.H.; Methodology, E.M.T.; Software, E.M.T. and A.C.J.R.; Validation, E.M.T. and M.M.; Investigation, E.M.T. and M.M.; Writing – Original Draft, E.M.T.; Writing – Review & Editing, E.M.T., C.E.V., A.C.J.R., M.M., M.I.B, and K.H.; Visualization, E.M.T.; Supervision, C.E.V., M.I.B, and K.H.

## DECLARATION OF INTERESTS

The authors declare no competing interests.

## SUPPLEMENTAL INFORMATION

### Cell-Level Model and Validation

We simulated simultaneous front and rear illumination in Synopsys TCAD Sentaurus using an optoelectronic model for crystalline silicon (c-Si) based encapsulated bifacial SHJ solar cells [S1], depicted in Figure S1. In our model, we use regular inverted pyramid texturing to emulate random pyramidal light-trapping. Optical calculations use Monte Carlo 3D ray tracing, with subwavelength hydrogenated amorphous silicon (a-Si:H) and indium tin oxide (ITO) thin films accommodated via transfer matrix method boundary conditions between the c-Si substrate and air. On both faces, we apply a 6.4% shading loss corresponding to the finger width and metallization coverage of the fabricated cells. Electrical calculations are solved using Poisson's equation and electron and hole continuity equations with the finite volume method in 2D. We apply surface recombination velocities between a-Si:H and c-Si layers of 2 cm/s [S2] and an ITO sheet resistance of 99.2 Ω/□, corresponding to our fabrication process [S3][S4]. The c-Si wafers used have a manufacturer-rated SRH lifetime of 2 ms, which we assume for our simulations. For a detailed description of our model and all other chosen parameters, see Ref. [S1].

We fabricated bare bifacial SHJ cells at Arizona State University for model validation. Crystalline silicon wafers were textured in a KOH bath before being cleaned in three 10-minute baths: a RCA-B (6:1:1 $H_2O$:$H_2O_2$:HCl) 74°C bath, a piranha (8:1 $H_2O$:$H_2SO_4$) 110°C bath, and a buffered oxide etch 25°C bath. Layers of doped and intrinsic hydrogenated amorphous silicon (a-Si:H) were deposited on cleaned 156×156 mm$^2$ c-Si wafers with an Octopus II (INDEOtec) plasma-enhanced chemical vapor deposition (PECVD) tool at 250°C. Indium tin oxide (ITO) was sputtered with a 2% oxygen gas flow to a thickness ~70 nm in an MRC 944 sputtering tool. Silver metallization was applied via screen printing on front and rear faces and cured/annealed for around half an hour at 200 °C.

Measurements of external quantum efficiency (EQE) and Suns-Voc under front-only and rear-only illumination were conducted to evaluate model performance as a function of wavelength and voltage, as given in Figure S1B and S1C. Measurements of EQE were conducted with an Oriel 300 W calibrated xenon lamp and Oriel Cornerstone 130 1/8 m Monochromator, while Suns-Voc measurements were conducted on a Sinton Instruments WTC-120.

Since EQE and Suns-Voc measurements were conducted on reflective chucks, a back-reflector was added to the simulations for this comparison, but it is not present for any other data analysis presented in this work. Under a spectrum of AM1.5G, modelled short-circuit current density ($J_{sc}$) has a discrepancy of 0.1 mA/cm$^2$ under front illumination and 0.3 mA/cm$^2$ in comparison to measured EQE-extracted $J_{sc}$. Thus, as inputs into Suns-Voc measurements, we apply pseudo short-circuit current densities equal to our simulation output. We provide these $J_{sc}$ input values in the first row of the table inset in Figure S1C. As Suns-Voc is an open-circuit voltage ($V_{oc}$) measurement technique, effects of series resistance are eliminated, resulting in the *pseudo J-V* curve presented in S1C by experimental dots. Modelled *J-V* behavior, including resistances, is given by solid lines for comparison.

Measured and modelled $V_{oc}$ shows excellent agreement, with a <5 mV discrepancy for both front and rear illumination. Pseudo-efficiency (*p-η*) output by Suns-Voc differs from our modelled efficiency (*η*) by 0.2% abs. and

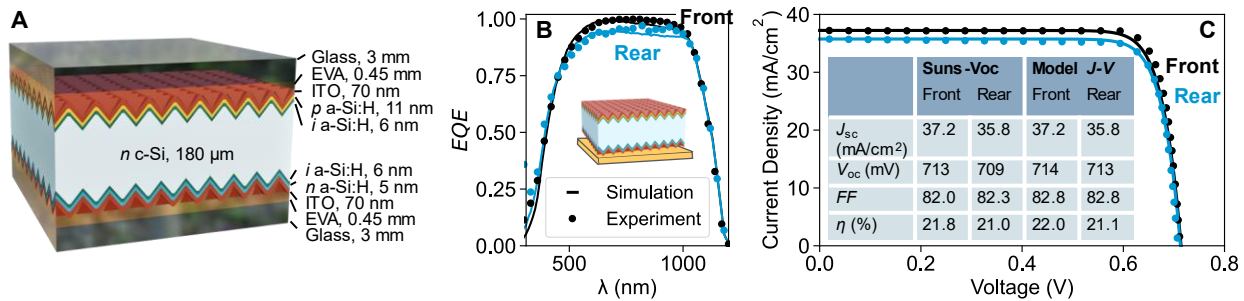

**Figure S1. Silicon Heterojunction Cell Model Validation**

(A) Encapsulated bifacial silicon heterojunction cell structure, as simulated in this work. Layer thicknesses are as labelled and align with cell fabrication. (B) Measured and simulated EQE for front-only and rear-only illumination of a bare cell on a reflective chuck. (C) Measured Suns-Voc with modelled *J-V*. Parameters extrapolated from Suns-Voc and modelled *J-V* curves are inset.

0.1% abs for front and rear illumination, respectively. Rear side $\eta$ is lower than front-side $\eta$ primarily due to lower $J_{sc}$, as depicted in S1B, caused by SRH recombination in the c-Si bulk. With higher c-Si wafer quality, reduced contamination during wafer cleaning, and optimized contacts, efficiencies around 24% are achievable [S5].

### The Scaled Rear Irradiance Method in Multiple Locations

The same analysis presented in the main paper has been repeated in Phoenix, USA and Ottawa, Canada to demonstrate variations in the calibration for other mid-latitude locations with different diffuse irradiance fractions. The results are summarized in Table S1 for the bifacial HSAT described in the main text. Ottawa, having a higher GHI-weighted ratio of diffuse light, has $R_{SRI}$ on average 3% higher than in Boulder. Phoenix, with more direct light, has on average 7% lower $R_{SRI}$. For either location, use of the calibrated values as given by Boulder results in similar energy yield deviations from full system-level analysis with DUET.

### Table S1. Calibration and Performance of the SRI Method in Multiple Locations

|  |  | Phoenix | Boulder | Ottawa |
|---|---|---|---|---|
| Latitude |  | 33.45°N | 40.01°N | 45.42°N |
| Annual DHI/GHI ratio |  | 0.27 | 0.36 | 0.43 |
| $R_{SRI}(x, \tau = HSAT)$ | Snow | - | 0.240 | 0.254 |
|  | White sand | 0.199 | 0.211 | 0.214 |
|  | Dry grass | 0.125 | 0.134 | 0.136 |
|  | Light sand | 0.095 | 0.102 | 0.104 |
|  | Concrete | 0.092 | 0.088 | 0.091 |
|  | Roof shingle | 0.080 | 0.086 | 0.088 |
|  | Green grass | 0.080 | 0.086 | 0.088 |
|  | Red brick | 0.062 | 0.067 | 0.070 |
|  | Soil | 0.054 | 0.059 | 0.062 |
| Annual predicted bifacial gain discrepancy with DUET using Boulder calibration values (% abs.) | IEC | 10 | 8 | 8 |
|  | Max | 27 | 33 | 32 |
|  | Scaled | 2 | 1.7 | 0.7 |

### The Spectral Scaled Rear Irradiance Method

As described in the main text, the scaled rear irradiance (SRI) method can be adapted to include the effects of spectral albedo via Equation 9 rather than a scaling of AM1.5G. We refer to this form of the SRI method as the spectral scaled rear irradiance (S-SRI) method. This method instead changes the applied front and rear spectra based on the spectral shape of the ground cover. Using the SRI method, output power of the full spectral, S-SRI calculation is predicted to within 0.2 W/m² on average. Efficiency for the SRI method accounts for the effect of spectral shape via Equation 7 and 8. Thus, efficiency calculated with S-SRI and SRI methods agree to within 0.03% abs. on average for the dataset of considered albedos.

### Additional Variations in Incident Spectra

Additional factors, such as atmospheric conditions and sun position, shift the incident spectrum received on bifacial cells and modules. This analysis could also be decomposed into direct and diffuse irradiance components. To keep calculations as simple as possible, and for ease of comparison with IEC 60904-1-2 standards, all calculations presented in the main text of this article have used the standard AM1.5G spectrum which assumes a particular ratio of direct-to-diffuse light and set values for atmospheric conditions. Under real-world conditions, this ratio depends on location, time of day, day of the year, system design, and other environmental factors [S6]. As diffuse light tends to have a higher UV content than direct light, locations with more diffuse light will have higher spectral-albedo enhancements for ground coverings that are UV-reflective, such as snow. For example, if the diffuse irradiance fraction increases by 10%, the ratio of EQE-weighted albedo to broadband albedo of snow will rise by around 1%.

## Summary of 3D View Factor System-Level Model Inputs

In Table S2 we summarize the environmental, cell, and system inputs into the 3D view factor model, DUET, for a bifacial horizontal single-axis tracked array. For fixed tilt array configurations, modules are South-facing at latitude-tilt. The performance of DUET has been compared to other software in Ref. [S7].

### Table S2. Summary of DUET Inputs

| | | |
|---|---|---|
| Environment | City | Boulder, Colorado, USA |
| | Latitude, longitude | 40.01°N, 105.26°W |
| | Sky model | Perez sky luminance distribution model |
| | Environmental data | NREL NSRDB |
| Cell geometry | Cell size | M2 (156.75 mm) |
| | Cell area (cm$^2$) | 238.84 |
| | Metallization shading fraction (%) | 7% |
| | Irradiance sample points per cell | 16 |
| Cell *I-V* | Bifaciality (%) | 96 |
| | Short-circuit current (A/cell) | 8.19 |
| | Responsivity (A/W) | 0.343 |
| | Open-circuit voltage (V) | 0.712 |
| | Ideality factor | 1.08 |
| | Saturation current (nA/cell) | 0.35 |
| | Shunt resistance (Ω) | 12.6 |
| | Series resistance (Ω) | 0.00018 |
| | Temp. coef. short-circuit current (K$^{-1}$) | 0.035 |
| | Temp. coef. open-circuit voltage (K$^{-1}$) | -0.00235 |
| | Nominal operating cell temp. (°C) | 42 |
| | Incidence angle modifier model | ASHRAE |
| Module geometry | Number of cells | 72 |
| | Module orientation | Portrait |
| | Frame? | Included |
| Array geometry | Row alignment | North-South |
| | Number of rows | 5 |
| | Row pitch (m) | 4.84 |
| | Number of tables per row | 1 |
| | Number of modules per table | 31 |
| | Module spacing (m) | 0.03 |
| | Number of tiers | 1 |
| | Ground clearance | 1.22 |
| | Racking? | Includes purlins and torque tube |
| Tracker | Type | East-West single-axis tracking |
| | Backtracking? | Included |
| Torque tube | Shape | Round |
| | Radius (m) | 0.05 |